\documentclass[11pt]{article} 

\usepackage{epigraph}
\usepackage{amsmath}
\usepackage{amssymb}
\usepackage{graphicx}
\usepackage{empheq}
\usepackage[algoruled,boxed,lined]{algorithm2e}

\usepackage[margin=3.0cm]{geometry}

\newcommand{\R}{\mathbb R}
\newcommand{\N}{\mathbb N}

\newcommand{\fer}[1]{(\ref{#1})}

\newcommand{\be}{\begin{equation}}
\newcommand{\ee}{\end{equation}}

\def\bqa{\begin{eqnarray}}
\def\eqa{\end{eqnarray}}

\def\e{\varepsilon}

\def\t{\theta}
\def\b{\beta}

\newcommand{\ff}{\widehat f}

\newcommand{\fg}{\widehat g}

\newcommand{\bd}{\begin{displaymath}}
\newcommand{\ed}{\end{displaymath}}
\newcommand{\ba}{\begin{eqnarray}}
\newcommand{\ea}{\end{eqnarray}}

\def\MM{\widehat M}
\def\GG{\widehat g}

\def\N{\mathbb{N}}
\def\R{\mathbb{R}}

\newcommand{\bq}{\begin{equation}}
\newcommand{\eq}{\end{equation}}

\newenvironment{equations}{\equation\aligned}{\endaligned\endequation}

\def\BigL{ {\mathcal {L}}}

\newtheorem{lemma}{Theorem}
\newtheorem{remark}{Remark}
\newcommand{\barr}{\begin{array}}
\newcommand{\earr}{\end{array}}

\def\q1{Q_\varepsilon^{(1)}}
\def\q0{Q_\varepsilon^{(0)}}

\begin{document}


\title{The kinetic theory of mutation rates}

\author{Lorenzo Pareschi
\thanks{Department of Mathematics and Informatics, University of Ferrara, Italy; lorenzo.pareschi@unife.it}\and 
Giuseppe Toscani\thanks{Department of Mathematics, University of Pavia,  Italy; giuseppe.toscani@unipv.it}\,\,\thanks{Institute of Applied Mathematics and Information Technologies,  Pavia Italy}
}



\maketitle

\abstract{The Luria--Delbr\"uck mutation model is a cornerstone of evolution theory and has been
mathematically formulated in a number of ways. In this paper we illustrate how this model of mutation rates can be derived by means of classical statistical mechanics tools, in particular by modeling the phenomenon resorting to methodologies borrowed from classical kinetic theory of rarefied gases.  The aim is to construct a linear kinetic model that can reproduce the Luria--Delbr\"uck distribution starting from the elementary interactions that qualitatively and quantitatively describe the variation of mutated cells.  The kinetic description is easily adaptable to different situations and makes it possible to clearly identify the differences between the elementary variations leading to the formulations of Luria--Delbr\"uck, Lea--Coulson, and Kendall, respectively. The kinetic approach additionally emphasizes basic principles which not only help
to unify existing results but also allow for useful extensions.}

\medskip
{\bf Keywords.} {Luria-Delbr\"uck distribution; kinetic theory; mutation rates; Fokker-Planck equations.} 

\section{Introduction}

 Statistical mechanics provides a powerful approach to the mathematical modeling of systems composed of a huge number of growing organisms interacting with each other and/or the environment, and has as its primary product the understanding of the relationship between parameters in microscopic rules and the resulting macroscopic statistical outcomes. In reason of this possibility, in the last two decades there has been a trend toward applications of statistical mechanics to interdisciplinary fields ranging from the classical biological context to new aspects of socio-economic phenomena \cite{CFL,NPT,PT13}.

In order to justify at best the statistical mechanics approach, classically used for gas dynamics where the number of molecules in a mole is quantified by Avogadro's constant \cite{Cer}, the system under consideration must possess some minimum requirements. On the one hand, the population of entities composing the system must be at least of the order of millions, and on the other hand, the time in which we observe the system must be high enough to ensure that we are close to a stable statistical profile. 
 
An issue that  naturally satisfies these requirements is the time-evaluation of  the number of mutation rates,  a challenging problem which has
attracted the interest of generations of both geneticists and mathematicians. This problem, which is essentially a matter  of evaluating the growth of a population, originated from a series of classical experiments
 conducted by Luria and Delbr\"uck \cite{LD}, which led to a fundamental example of how mathematics can be used to estimate mutation rates. The model proposed by Luria and Delbr\"uck assumed deterministic growth of mutant cells, which seemed too stringent an assumption to allow for efficient extraction of reliable information about mutation rates from experimental data.

This shortcoming of the Luria and Delbr\"uck model was, some years later, remedied by a slightly different mathematical formulation proposed by Lea and Coulson \cite{LC}, who adopted the Yule stochastic birth process to mimic the growth of mutant cells. In the ensuing decades the Lea--Coulson \index{cell mutation
models!Lea--Coulson model} formulation occupied so prominent a place in the study of mutation rates that the Lea--Coulson formulation is now commonly referred to as the Luria--Delbr\"uck model, as documented by the long list of contributions to the study of the Lea--Coulson formulation, including
\cite{Ang,Arm,Bar,CH,Jon,Koc,Ma,Man,Pak}, and numerous others cited in the essential review article by Zheng \cite{Zhe0}, which appeared at the end of the last century. In his paper, Zheng examines the various mathematical formulations, focusing on important practical issues closely linked with the distribution of
the number of mutants. It appears evident that almost all of these formulations are prominently based on probabilistic arguments.

Adopting the point of view of statistical mechanics, the Luria--Delbr\"uck problem has been first modeled only few years ago by Kashdan and one of the present authors in \cite{KP}. There, a kinetic description of the density of mutation rates has been considered, in which the Luria--Delbr\"uck distribution appears as a suitable limit of a linear kinetic equation describing the growth of mutated cells. Following Luria and Delbr\"uck's original model, mutation is, in the kinetic picture, a consequence of a specific
\emph{updating elementary rule}, under which both the normal cells and the mutant cells grow deterministically but mutations occur randomly. The analysis in \cite{KP} has been subsequently extended by one of the present authors in \cite{To13}  to cover the Lea and Coulson model \cite{LC}, which is based on a different formulation under which the normal cells grow deterministically but the mutants grow stochastically \cite{Ken}.

As discussed in \cite{Sak},  the kinetic equations introduced in \cite{KP, To13} (see also \cite{PT13}) are well-posed from a mathematical point of view, and various properties of the unique solutions are available to be studied from a numerical point of view. 

The pioneering work \cite{KP} was credited with outlining the power of kinetic methodology to model growth processes in the field of mutation rates.

Indeed, for years various concepts and techniques of statistical mechanics have been successfully applied to a wide variety of complex systems, physical and otherwise, in an attempt to understand, starting from the elementary interactions, the emergent properties that appear therein.

 In particular, this methodology has been employed to write down realistic evolution models
for wealth in economics \cite{BPTZ, Ch02, ChaCha00,ChChSt05, CoPaTo05, DPTZ, DY00,Ha02,Igle, Sl04}. In this context, it was rigorously shown that the asymptotic
profile of wealth distribution, and the possibility to have Pareto-type tails, depends completely on the microscopic structure of binary trades \cite{DMT, MaTo07}.

In the kinetic description of mutation rates, the core of the methodology is to identify the linear \emph{elementary interaction rule} which drives the underlying process of mutation, in order to write a kinetic equation of the Boltzmann type for the distribution
density of the mutated cells. As is common in this approach,  the construction of the kinetic equation emphasizes the differences between the microscopic interaction rules of the original
Luria--Delbr\"uck distribution and the Lea--Coulson modification.
Similarly to the collisional Boltzmann equation of rarefied gases in the case of Maxwell-type interactions \cite{Cer}, the kinetic description allows us to use in a natural and powerful way the weak formulation in terms of the Fourier transform \cite{Bob88}, which clarifies the nature of the assumption made by Lea and Coulson \cite{LC}, namely the choice of a Yule process for the random growth of the mutated cells \cite{Ken}. 

This challenging application shows that the kinetic description based on linear interactions is powerful and flexible, and
allows us in most cases to establish a direct connection between the
\emph{elementary interaction} and the consequent \emph{macroscopic
behavior}. Once again, as happens in the study of wealth distribution in a multi-agent society, where the
characteristics of the binary trade reflect into the steady profile
of wealth, or in opinion formation \cite{To06}, where the details of the binary
exchange of opinion produce the steady distribution profile of opinion in the
system, one can learn from microscopic dynamics the forthcoming
macroscopic behavior of mutant cells, and associate with the
various models the corresponding elementary mechanism of interaction.

Moreover, the kinetic description in terms of elementary interactions indicates in most cases the right way to design numerical algorithms useful to follow the process under study, where the numerical solution satisfies the main physical properties of the continuous one \cite{PZ, DPZ}. Consequently, it gives a new possibility to approach the numerical simulation problem of mutation rates, and to compare it with existing ones \cite{Zhe1, Zhe2}.


\section{Basics of kinetic theory}\label{sec:Galton}

The objective of linear kinetic theory is to determine the time evolution of the statistical distribution of a phenomenon subject to frequent elementary variations of the same type.  Let $x \in I \subseteq \R$ denote the amount of a certain evolving quantity (wealth, opinion, weight, $\dots$), and let us denote by $f(x,t)\, dx$, $x \in \R$,  the amount of this quantity which at time $t\ge 0$  lies in the  interval $(x, x +dx)$. Suppose moreover that the quantity $f(x,t)$   changes over time due to repeated linear elementary variations given by
  \be\label{ele-gen}
  x_* = Ax + B,
  \ee
  where the coefficients $A$ and $B$ can be constant or, more generally, random quantities. According to classical kinetic theory \cite{Cer, PT13}, the  evolution in time of the density $f(x,t)$  is captured by looking at its time variation  along the \emph{observable quantities} $\varphi(x)$, given by the equation
\be\label{bo}
\frac d{dt} \int_\R \varphi(x)f(x,t)\, dx = \frac 1\tau \left\langle \int_\R  \left[ \varphi(x_*) -\varphi(x)\right] f(x,t) \,dx\right\rangle.
 \ee
The integral on the right-hand side of equation \fer{bo} measures the variation of the observable quantity $\varphi$ consequent to the variation of the quantity $x$ under consideration, determined by the elementary law \fer{ele-gen}. The presence of the mean $\langle\cdot\rangle$ is due to the (possible) presence of  random quantities in  the coefficients $A$ and $B$. The positive part of the integral quantifies the variation of $\varphi$ consequent to the change  from $x_*$ to $x$ (gain term),  while the negative part of the integral quantifies the variation of $\varphi$ consequent to the change  from $x$ to $x_*$ (loss term). Last, the time frequency $1/ \tau$ is a measure of the speed at which the phenomenon relaxes towards the steady or self-similar solution \cite{Cer}. This quantity can be easily absorbed by a suitable scaling of the time variable.

Note that, starting from equation \fer{bo}, it is immediate to determine the evolution of the principal moments of the distribution function $f(x,t)$, that are obtained by choosing the observable quantity $\varphi(x)= x^n$, $0\le n \in \N$, and consequently to identify eventual conservation laws. 

As recently described in \cite{GTV},  this methodology can be profitably applied both to describe the well-known Galton's experiment  \cite{Gal1,Gal2}, and to obtain information  about its behavior over long periods of time.  Indeed, Galton bean machine represents a clear visual approach to link  {repeated elementary interactions} to a consequent {universal} steady state density.  In this famous experiment, balls subject to gravity exhibit repeated symmetric variations of direction before falling on the ground. Each variation is determined by interaction of the ball with the vertex of an isosceles triangle.  At the bottom, if {the number of balls is sufficiently high}, one can catch a glimpse of the profile  of the  {Gaussian} density 
 \[
 M(x) = \frac 1{\sqrt{2\pi\sigma}}\exp\left\{-\frac{x^2}{2\sigma} \right\}.
 \]
Galton experiment is equivalent to a sequence of \emph{Bernoulli trials} (or binomial trials), in which the trial is a random experiment with exactly two possible outcomes, {success} and {failure}. 
  The probability of success ({$1/2$} in Galton experiment) is the same every time the experiment is conducted. 
 The classical way to give a mathematical interpretation of the result is to pass from the discrete to the continuous description. In the half-plane $(x,t)$, $t\ge 0$  the {continuous} description of the time variation of the probability  $p(x,t)dx$  to find the ball in the interval $(x, x+\Delta x)$ at time $t +\Delta t \ge 0$ is expressed by the relationship
  \[
  p(x,t+\Delta t) = \frac 12 p(x+ \Delta x, t ) + \frac 12 p(x- \Delta x, t ).
  \]
Expanding in Taylor's series, at the first order in time the left-hand side and at the second order in position the right-hand side, shows that in the limit {$(\Delta x)^2/\Delta t \to \sigma>0$}, $p(x,t)$ satisfies the linear diffusion equation
 \[
 \frac{\partial p}{\partial t} = {\frac\sigma{2} }\frac{\partial^2 p}{\partial x^2}.
 \]
At this point, it is enough to observe that in Galton experiment the repeated trials start from $x=0$.  Consequently, if the number of repeated trials is sufficiently large, the resulting profile is well approximated by the \emph{source-type solution} of the linear diffusion equation.
   The {source-type solution} to the diffusion equation at time $t >0$ is the {Gaussian density}
 \[
 M(x, t) = \frac 1{\sqrt{2\pi\sigma t}}\exp\left\{-\frac{x^2}{2\sigma t} \right\}.
 \]
This gives an answer to the visual result of the experiment.

Let us now resort to the classical approach of collisional kinetic theory described above \cite{GTV}. Let us consider a huge number of identical Galton bean machines, and let us run a trial of the experiment simultaneously on all machines.  Let us denote by $f(x,t)\, dx$, $x \in \R$,  the percentage of balls which at time $t\ge 0$  lie in the  interval $(x, x +dx)$. The quantity $f(x,t)$ changes in time since balls are subject to interactions which modify their position.  The {elementary interaction} of type \fer{ele-gen} is now given by
  \be\label{ga-in}
  x_* = x + \sqrt\sigma \eta,
  \ee
 where $\eta$ is a {symmetric random variable} taking values $\pm 1$ with probability $1/2$ ( $2\sqrt\sigma$  is the length of the base of  the triangles). 
Let us further suppose that the trial starts with balls located in $x=0$, and let us compute, via the kinetic equation \fer{bo},  the time evolution of the principal moments.

The choice $\varphi(x) = 1$ gives
  \[ 
  \frac d{dt} \int_\R   \,f(x,t)\, dx =  0,
  \]
which shows that the mass is preserved in time. This means that $f(x,t)$ remains a probability density for all $t \ge 0$, if it is so initially.

Analogously,   if we choose $\varphi(x) = x$, since {$\langle x_* -x \rangle = \langle \eta \rangle = 0$}, we find that the mean value is preserved in time.
Thus, the mean value of $f(x,t)$, equal to zero at time $t=0$, is {equal to zero for all times $t \ge 0$}.
 
Moreover, since {$\langle x_*^2- x^2\rangle = \langle 2x\sqrt\sigma\eta + \sigma\eta^2\rangle = \sigma$},  
\[
\frac d{dt} \int_\R  x^2 f(x,t)\, dx =  \frac\sigma\tau  \int_\R  f(x,t)\, dx = \frac\sigma\tau.
 \]
 Consequently, the variance of the density function {diverges with time}. This shows that we can not expect that the solution will tend for large times towards a  steady state of finite variance. 
 
 This obstacle can be circumvented by suitably {shrinking the domain}. Given a positive constant $\lambda <1$ let us consider the modified interaction 
  \be\label{ga-sc}
 x_* = x(1-\lambda) + \sqrt\sigma\eta,
  \ee
  which still belongs to the class of interactions \fer{ele-gen}.
  
The shrinking does not modify the characteristics of the shape of $f(x,t)$. Also, the mean value of $f(x,t)$  remains {equal to zero for all times $t \ge 0$}.
 However, now {$\langle x_* ^2 -x^2\rangle = \sigma - \lambda(2-\lambda)x^2$}. Consequently
 \be\label{vava}
\frac d{dt} \int_\R  x^2 f(x,t)\, dx =  \frac\sigma\tau \int_\R   f(x,t)\, dx  - \frac{\lambda(2-\lambda) }\tau \int_\R  x^2 f(x,t)\, dx.
 \ee
The variance of the density function {remains uniformly bounded in time}. Indeed, solving the differential equation \fer{vava} we obtain
  \[
  \int_\R  x^2 f(x,t)\, dx =  e^{-\lambda(2-\lambda)/\tau} \,\int_\R  x^2 f_0(x)\, dx  + {\frac{\sigma}{\lambda(2-\lambda)}} \left[ 1- e^{-\lambda(2-\lambda)/\tau} \right].
\]
This suggest that, in presence of the elementary variation \fer{ga-sc}, the solution to the kinetic equation could relax towards a \emph{universal} steady state. This characteristic can be captured by resorting to a second ingredient, which is now part of classical kinetic theory, which consists in studying the problem in the so-called \emph{grazing collision limit} (cfr. \cite{FPTT} and the references therein). By this limit procedure, we mainly consider only interactions which produce  a very small modification of the  pre-interaction value (from this the name \emph{grazing} which is borrowed from the kinetic theory of rarefied gases, where grazing interactions describe a binary collision in which the outgoing post-collision velocities remain very close to the ingoing ones), while waiting enough time to see a finite variation of the observable quantities.    
  
Let us apply the grazing procedure to the modified Galton experiment.   
We {reduce the size} of the triangles, the shrinking parameter and the time frequency in such a way to still observe the evolution of the variance towards a {value bounded away from zero}. By looking at the coefficients of equation \fer{vava} we realize that this can be done by setting, for $\e \ll 1$
 \be\label{scala}
 {\sigma \to \e \sigma; \quad \lambda \to \e \lambda; \quad \tau \to \e\tau}.
 \ee
The meaning of the scaling \fer{scala} is clear.   Within this scaling, each {interaction produces only a small variation} of the position variable. To see its effect on the variation of the observable, one has to suitably increase the time frequency of the interactions!

In this regime, for a given regular observable function $\varphi(x)$, expanding in Taylor series up to the second order, and neglecting terms of order $o(\e)$ (cf. \cite{FPTT} for details) we can write
  \[
  \langle \varphi(x_*) - \varphi(x)\rangle \simeq \varphi'(x) \langle x_* -x \rangle + \frac 12 \varphi''(x) \langle( x_* -x )^2\rangle \simeq
  \]
  \[
 {\e }\left[ - \lambda \, x \, \varphi'(x) + \frac \sigma 2\,   \varphi''(x) \right].  
  \]
  
 Setting for simplicity $\tau = {\e} $ we conclude that the kinetic equation is well approximated by
\be\label{fp}
\frac d{dt} \int_\R \varphi(x)f(x,t)\, dx = \int_\R \left[{ - \lambda\, x\, \varphi'(x) + \frac \sigma 2\,   \varphi''(x)} \right] f(x,t) \,dx.
 \ee
Equation \fer{fp} is the weak form of the Fokker--Planck equation \cite{FPTT}
\[
\frac{\partial f(x,t)}{\partial t} ={\frac\sigma 2}  \frac{\partial^2 f(x,t)}{\partial x^2} + {\lambda}\frac{\partial(x\, f(x,t))}{\partial x}.
 \]
The equilibrium solution of the Fokker--Planck equation (of unit mass) is the {Gaussian probability density } \cite{FPTT}
\be\label{gd}
 f_\infty(x) = \sqrt{\frac {{\lambda}} {{4\pi \sigma}}} \exp \left\{ -\frac{\lambda x^2}\sigma\right\}.
 \ee
The kinetic approach justifies the {experimental evidence}. 

\begin{remark}
The kinetic description remains valid even in the case in which the shrinking parameter $\lambda$ is assumed equal to zero. In this case, instead of the Fokker--Planck equation we obtain that the grazing limit produces the linear diffusion equation classically obtained by adopting the continuous description in the half-plane we briefly illustrated before.
\end{remark}

\begin{remark}
At difference with the classical Galton bean machine, where, as explained above,  the normal density can be observed only if all the repeated trials start from $x=0$, in the kinetic approach, that leads to the Fokker--Planck equation, any initial probability density  of finite variance  is shown to converge towards the Gaussian density \fer{gd} \cite{FPTT}.
\end{remark}

\section{The kinetic description of mutation rates}\label{sec:mutation}

The procedure of Section \ref{sec:Galton} is quite general, and can be easily adapted for studying the problem of mutation rates. This will be the object of this Section. Before illustrating the kinetic approach, let us briefly discuss the classical setting. 

The most-used formulations of a mutation are generally expressed by 
\begin{itemize}
\item[(A)] grey$\to$ black, 
\item[(B)] grey$\to$ grey $+$ black.
\end{itemize} 
Here, \emph{grey}
stands for a normal cell and \emph{black} a mutant cell. Note that
under formulation (A), a mutation entails the loss of the normal
cell. In formulation (B), the normal cell continues to live after a
mutation. Consequently, the occurrence of a mutation does not
decrease the rate at which mutations occur in subsequent time
intervals. Assumption (B) was implicitly made by Lea and Coulson
\cite{LC} and Armitage \cite{Arm}. Since our goal is to present a
kinetic description of the Lea--Coulson formulation, in what follows
we will assume formulation (B).

The first mathematical description of the mutation process, which was
subsequently modified by Lea and Coulson, goes back to Luria and
Delbr\"uck \cite{LD}. Because all subsequent formulations are simple
variations of this formulation, we detail below its underlying
assumptions.

The process starts at time $t=0$ with one normal cell and no
mutants. Normal cells are assumed to grow deterministically at a
constant rate $\b_1$. Therefore, the number of normal cells at time
 $t>0$  is $N(t)= e^{\b_1 t}$. 
 
 Mutants grow deterministically at a constant rate $\b_2$. If a
mutant is generated by a normal cell at time $s> 0$, then the clone
spawned by this mutant will be of size $e^{\b_2(t-s)}$ for any
$t>s$.  Mutations occur randomly at a rate proportional to $N(t)$.
If $\mu$ denotes the per-cell per-unit-time mutation rate, then the
standard assumption is that mutations occur in accordance with a
Poisson process having the intensity function
 \be\label{8me}
\nu(t) = \mu N(t)= \mu e^{\b_1 t}.
 \ee
Consequently, the expected number of mutations occurring in the time
interval $[0,t)$ is
 \[
\int_0^t \nu(s) \, ds = \frac \mu{\b_1}\left( e^{\b_1 t} -1 \right).
 \]
Let $X(t)$ denote the number of mutants existing at time $t$, and $f(n,t)$
the probability of having $n$ mutants at time $t$. The previous
assumptions imply that $X(t)$ can be expressed as
 \be\label{8LL}
X(t) = \sum_{i =1}^{M(t)}\exp\{\b_2(t -\tau_i)\}, \quad {\rm if} \quad M(t)
\ge 1,
 \ee
while
 \[
X(t) = 0 \quad {\rm if} \quad M(t) = 0\, .
 \]
Here, $\tau_i$ are the random times at which mutations occur, and
$M(t)$ stands for the mutation process which is a Poisson process
with intensity function $\nu(t)$ given in \fer{8me}. The
probabilistic description \fer{8LL} allows for recovering the
various quantities like mean, variance and cumulants (cf. the
detailed historical description in \cite{Zhe0}).

An interesting alternative to the probabilistic construction
described above, is to use the approach of Section \ref{sec:Galton}. Within this approach,
the study of the time evolution of the probability distribution
$f(n,t)$ of mutants, together with a reasonable explanation of the
growth process induced by this distribution, is achieved by means of linear
kinetic models \cite{KP, To13}. 

In the rest of the paper, we will assume for simplicity a
continuous parameter $v$ for the number of mutants, and that the time
variation of the distribution $f(v,t)$ of mutants is based on elementary interactions of type \fer{ele-gen}.
The interaction rule described by the previous probabilistic
description indeed corresponds  to assuming that, at time $t
>0$, the microscopic growth of mutants is driven by the elementary law of variation
\begin{equation}
  \label{8colli1}
  v^*  = (1 + \beta_2) v  + \eta(t),
\end{equation}
where $\eta \ge 0$ (absence of backward mutations) is a discrete
random  variable
distributed according to a Poisson density $p(\nu(t))$ of intensity
given by \fer{8me}. 

This makes an essential difference between the elementary interactions considered in Section \ref{sec:Galton} 
and the present one. Now, the intensity of the elementary interaction varies in time, due to time variable random contribution. In reason of this, the principal moments of the distribution, including the mean value, are time dependent.

Also, for any given $v$, the number of mutants
after the interaction can only assume the values
 \[
v_i^*  = (1+ \beta_2) v  + i, \quad i \in \N.
 \]
The interaction \fer{8colli1} then leads to the kinetic equation (in weak \index{cell mutation models!kinetic equation}
form) \cite{KP}
 \be\label{8ki}
\frac{d }{dt}\int_{\R_+}f(v,t)\varphi(v)\,dv = \sum_{i=1}^\infty
p_i(\nu(t))\int_{\R_+}  \bigl( \varphi(v_i^*)-\varphi(v)\bigr)
f(v,t) \,dv,
 \ee
where $\varphi(v)$ is a smooth function. The condition of having no
mutants at time $t=0$ translates into the initial condition
 \be\label{8in}
 f(v, t=0) =f_0(v) = \delta_0(v),
 \ee
where $\delta_a( v )$ indicates as usual the Dirac delta function
concentrated at $v=a$.

Interaction \fer{8colli1} belongs to the class of linear interactions \fer{ele-gen},
and it is sufficiently flexible to describe both the original
Luria--Delbr\"uck and Lea--Coulson formulations.

In any of the aforementioned formulations of the growth of the
mutant cells, the microscopic process is due to two simultaneous
events. The first one describes the intrinsic growth of the mutants,
which can be represented by a general interaction of type
 \be \label{8inter}
v' = v+ \BigL(v),
 \ee
where $v \to \BigL(v) \ge 0$,  as in \fer{ele-gen}, is a linear map acting on $v$. Note that the
choice \index{cell mutation models!Luria--Delbr\"uck model}
$\BigL(v) = \beta_2v$ gives the classical Luria--Delbr\"uck relation
\fer{8colli1}. The map  can be either deterministic or random. In
the latter case, corresponding to the Lea--Coulson formulation,
denoting by $\langle \cdot\rangle$ the mathematical expectation, in
order to maintain the same mean growth of the deterministic
interaction we will assume that
 \be\label{8gro}
 \langle \BigL(v) \rangle =  \beta_2 v.
 \ee
The second event describes the growth of the mutant cells due to the
presence of the normal ones which can mutate according to assumption
(B). In our picture this corresponds to assuming that the mutants
are immersed in a background (of normal cells), which, due to their
random mutation rule, undergo interactions which increase the number
of mutants according to the time-dependent probability density
$M(w,t)$. In the Luria--Delbr\"uck setting the probability
density of the background depends on time, and it is a probability mass function of Poisson type, with intensity $\nu(t) = \mu N(t)$. 

This corresponds to say that the background density takes the form
\[
M(w,t) = \sum_{k \ge 0} \frac{\nu(t)^k e^{-\lambda}}{k!} \delta_k (w),
\]
where, as  in \fer{8in},  $\delta_k (w)$ indicate the Dirac delta functions located in the points $k \in \N_+$.
Consequently,
 \be\label{8cond-b}
\int_{\R_+}M(w,t)\,dw = 1 \, ; \quad \int_{\R_+}w M(w,t)\,dw = \mu e^{\b_1
t}.
 \ee
Coupling the interaction \fer{8colli1} with the contribution due to the
background, one obtains the general elementary interaction
\begin{equation}
  \label{8colli}
  v^*  = v + \BigL(v) + w.
\end{equation}

Following the arguments of Section \ref{sec:Galton},  the effect of
interactions \fer{8colli} on the time variation of the mutant
density can be quantitatively described by a linear Boltzmann-type
equation in which the variation of the density $f=f(v,t)$ is due to
a balance between gain and loss terms that, for the given value
$v$, take into account all the interactions of type \fer{8colli}
which end up with the number $v$ (gain term) as well as all the
interactions which, starting from the number $v$, lose this value
after interaction (loss term). For a given observable $\varphi(v)$, the kinetic equation for the density $f(v,t)$ is given by
 \be
  \label{8kine-w}
\frac{d }{dt}\int_{\R_+} \varphi(v) f(v,t)\,dv = \left\langle
\int_{\R_+^2} \bigl( \varphi(v^*)-\varphi(v)\bigr) f(v,t) M(w,t)
\,dv \, dw \right\rangle.
 \ee
In equation \fer{8kine-w}, the presence of $\langle \cdot\rangle$ takes into
account the possibility that $\BigL(v)$ could be a random quantity.
Note that the choice $\varphi(v) =1$ implies that $f(v,t)$ remains a probability
density if it is so initially:
 \be\label{8m0}
\int_{\R_+} f(v, t)\,dv = \int_{\R_+} f_0(v)\,dv = 1 .
 \ee
Moreover, given the elementary interaction \fer{8colli}, choosing $\varphi(v) = v$ shows
that the total mean number of mutants satisfies the differential equation
 \be\label{8mm0}
 \frac{dm(t)}{dt} = \int_{\R_+}\langle \BigL(v)\rangle f(v, t)\,dv +  \mu e^{\b_1 t}.
 \ee
We note that the choice $\langle \BigL(v)\rangle = \b_2v$ implies
 \be\label{8mm}
 \frac{dm(t)}{dt} = \b_2 m(t) +  \mu e^{\b_1 t},
 \ee
which, coupled with the initial condition $m(0) = 0$ induced by
\fer{8in}, leads to the explicit computation of the mean value of
the Luria--Delbr\"uck distribution ($\b_1 \not= \b_2$), for any time
$t>0$,
 \[
m(t) = \frac \mu{\b_1 -\b_2}\left(e^{\b_1 t} - e^{\b_2 t}\right).
 \]
This expression is independent of the particular form of the map
$\BigL$, provided the map is such that condition \fer{8gro} is
satisfied. In other words, at the level of the growth of the mean
value, the original formulation by Luria--Delbr\"uck, and its
modification considered by Lea--Coulson, produce the same time
behaviour.

\subsection{The role of Fourier transform} \label{sec:Fourier}

As is usually done for this type of kinetic model, most of the
analytical properties of the solution can be obtained by resorting to
Fourier analysis. The choice $\varphi(v) = e^{-i\xi v}$ in the  formulation \fer{8kine-w}  is, in fact,
equivalent to the Fourier-transformed equation
 \begin{equation}
  \label{8eq}
  \frac{\partial \widehat f(\xi, t)}{\partial t}
  = \MM(\xi,t)\int_{\R_+}\langle e^{-i\xi\BigL(v)}\rangle e^{-i\xi v} f(v, t)\,dv  - \ff(\xi,
  t).
 \end{equation}
In \fer{8eq} we used the fact that $M(w,t)$ is a probability mass
function, so that $\MM(0, t)= 1$.  The initial condition \fer{8in} turns
into
\[
\widehat{f}_0(\xi)=1.
\]
Note that, since $M(w,t)$ is a probability mass function of Poisson type with intensity $\nu(t) = \mu
e^{\b_1 t}$,
 \[
\MM(\xi, t) = \exp\left\{\mu e^{\b_1 t}\left(e^{-i\xi} -1 \right)
\right\}.
 \]
Clearly, the explicit evaluation of the gain term on the right-hand side of
equation \fer{8eq} requires the knowledge of $\BigL(v)$. In the case
in which $\BigL(v) = \b_2 v$, equation \fer{8eq} takes the form \cite{PT13, To13}
 \begin{equation}
  \label{8eqLD}
  \frac{\partial \widehat f(\xi, t)}{\partial t}
  = \MM(\xi,t)\ff((1+\b_2)\xi)  - \ff(\xi, t).
 \end{equation}

Likewise, by assuming that $\BigL(v)$ is a Poisson variable of mean value
$\b_2 v$, thus satisfying \fer{8gro},
 \be\label{8pl}
\langle e^{-i\xi\BigL(v)}\rangle =  \exp\left\{ \b_2 v\left( e^{-i\xi} -1
\right) \right\} = \exp\left\{-iv \b_2 i \left( e^{-i\xi} -1 \right)
\right\},
 \ee
and equation \fer{8eq} becomes \cite{To13}
\begin{equation}
  \label{8eqLC}
  \frac{\partial \widehat f(\xi, t)}{\partial t}
  = \MM(\xi,t)\ff\left( \xi + \b_2 i \left(
e^{-i\xi} -1 \right), t \right)  - \ff(\xi, t).
 \end{equation}

Equations \fer{8eqLD} and \fer{8eqLC} differ only in the gain term on the
right-hand side, where the Fourier transform of the probability density is
evaluated at different points.

The existence of a solution to equation \fer{8eqLD} and \fer{8eqLC}
can be seen easily using  well-consolidated techniques of common use in kinetic theory of social phenomena \cite{FPTT, Sak}. 
The main argument  is to resort to the class of Fourier-based metrics introduced in \cite{GTW95}. 

 For a given pair of random variables $X$ and $Y$ distributed according to $f$ and $g$ these metrics read
\be\label{me-inf}
 d_r(X,Y)  = \sup_{\xi\in\R} \frac{|\ff(\xi) -\fg(\xi)|}{|\xi|^r}, \quad r >0.
 \ee
As shown in \cite{GTW95}, the metric $d_r(F,G)$ is finite any time the probability measures $f$ and $g$ have equal moments up to $[r]$, namely the entire part of $r\in \R_+$, or equal moments up to $r-1$ if $r \in \N$, and it is equivalent to the weak$\null^*$ convergence of measures for all $r >0$. Among other properties, it is easy to see \cite{GTW95, PT13} that, for two pairs of  random variables  $X,Y$, where $X$ is independent from $Y$, and  $Z, \tilde Z$ ($Z$ independent from $\tilde Z$),  and any  constant $c$ 
 \begin{equations}\label{sca1}
 & d_r(X+Y, Z+\tilde Z)\le d_r(X,Z) + d_r(Y,\tilde Z) \\
 & d_r(c X, c Y) =  |c|^r d_r(X,Y).
 \end{equations}
These properties classify $d_s$ as an ideal probability metric in the sense of Zolotarev \cite{Zolo}. 

To illustrate how these metric work, we give some details. 

Suppose the goal
is to prove that the operator on the right-hand side of \fer{8eq} is
Lipschitz continuous in this metric. 
 In order to show this, if $\ff_1(\xi, t)$ and $\ff_2(\xi,t)$ are two
solutions to equation \fer{8eqLD} corresponding to the initial data
$\ff_{0,1}(\xi)$ and $\ff_{0,2}(\xi)$, it holds that
 \begin{eqnarray*}
&& \frac{\partial }{\partial t}\frac{\ff_1(\xi,t) - \ff_2(\xi,t)}{|\xi|^s} +  \frac{\ff_1(\xi,t) - \ff_2(\xi,t)}{|\xi|^s}\\
&&\quad = \frac {\ff_1((1+\b_2)\xi, t) - \ff_2((1+\b_2)\xi, t)}{|\xi|^s}\MM(\xi,t).
 \end{eqnarray*}
On the other hand, since $|\MM(\xi,t)| \le 1$,
 \bqa
 \nonumber
&&\sup_{\xi \in \R} \left\vert \frac {\ff_1((1+\b_2)\xi, t) -
\ff_2((1+\b_2)\xi, t)}{|\xi|^s}\MM(\xi,t)\right\vert\\[-.25cm]
\label{8low}
\\[-.25cm]
\nonumber
&&\quad\le \sup_\xi \left\vert \frac {\ff_1((1+\b_2)\xi, t) - \ff_2((1+\b_2)\xi,
t)}{|\xi|^s}\right\vert = (1+\b_2)^s d_s(f_1(t), f_2(t)).
 \eqa
Thanks to \fer{8low},
 \[
\left\vert \frac{\partial }{\partial t}\frac{\ff_1(\xi,t) -
\ff_2(\xi,t)}{|\xi|^s} +  \frac{\ff_1(\xi,t) -
\ff_2(\xi,t)}{|\xi|^s}\right\vert \le (1+\b_2)^s d_s(f_1(t), f_2(t)),
 \]
 which, by Gronwall's inequality, implies
 \be\label{8decay}
 d_s(f_1(t), f_2(t)) \le d_s(f_{0,1}, f_{0,2})\exp \left\{\left((1+\b_2)^s - 1)\right)t
 \right\},
 \ee
which implies uniqueness.

The same method applies to equation \fer{8eqLC}. In this case, by choosing
$s = 1$, formula \fer{8low} reads
 \bqa
 \nonumber
&&\sup_\xi \left\vert \frac {\ff_1\left( \xi +\b_2 i \left( e^{-i\xi} -1
\right), t\right) - \ff_2\left( \xi + \b_2 i \left( e^{-i\xi} -1 \right),
t\right)}{|\xi|}\right\vert|\MM(\xi,t)|\\[-.25cm]
\label{8low2}
\\[-.25cm] 
\nonumber
&&\quad \le \sup_\eta \left\vert \frac {\ff_1\left(\eta, t\right) -
\ff_2\left(\eta, t\right)}{|\eta|}\right\vert
 \frac{\vert(\xi + \b_2 i \left( e^{-i\xi}
-1\right)\vert}{|i\xi|} \le (1+\b_2) d_1(f_1(t), f_2(t)).
 \eqa
The previous analysis indicates that one can obtain both existence and
uniqueness of a solution to equation \fer{8eq} for a wide choice of the
map $\BigL(v)$.

The Fourier-transformed kinetic equation further allows for a direct
evaluation of \index{cell mutation models!cumulant generating
function} the evolution equation for the cumulant-generating
function, defined as
 \be\label{8cum}
k(\xi,t) = \log \ff(\xi,t).
 \ee
As an example, a simple computation on equation \fer{8eqLD} shows
that $k(\xi,t)$ solves
 \be\label{8cum2}
\frac{\partial k(\xi, t)}{\partial t}
  = \exp\left\{ k((1+\b_2)\xi, t)  - k(\xi, t) \right\} \MM(\xi,t) -1.
 \ee
Generally, the cumulants can be extracted from the cumulant-generating
function via differentiation (at zero) of $k(\xi, t)$. That is, the
cumulants appear as the coefficients in the Maclaurin series of $k(\xi,
t)$. In the case under consideration,  the modulus of the first
coefficient coincides with the mean value, and the second one with the
variance. The evaluation of the subsequent derivatives of $k(\xi, t)$
appears extremely cumbersome, and the best one can do explicitly is to
evaluate the first two coefficients of the Maclaurin series. On the other
hand, since the explicit expression of the solution to the Boltzmann
equation \fer{8eq} is not known, the description of the solution though
its cumulants is one of the main operative possibilities.

\section{The quasi-invariant limit of the growth of mutant
cells}\label{8inva}

As it happens in the Galton experiment discussed in Section \ref{sec:Galton},  in general it is quite difficult to
obtain explicit results, or to describe precisely the main properties of the
solution to equation \fer{8eq}.  Also, as explained at the end of the
previous section, it is cumbersome to deduce from equation
\fer{8cum2} a recursive relation which allows for  detailed
computation of the cumulants. This difficulty is well known to
people working on this topic, where explicit results are rarely
present, independent of the different formulations. 

From this point
of view, while alternative to standard probabilistic approaches, the
kinetic picture does not produce evident simplifications.

On the other hand, as discussed in Section \ref{sec:Galton}, the kinetic formulation allows for the possibility to make use of the \emph{grazing asymptotics}, which, as we showed in the case of Galton experiment,  result in simplified models  of
Fokker--Planck type, for which it is relatively easier to achieve in
many cases the relevant properties of the solution, if not its
analytic expression.

In order to give, in the present case, a physical basis to these
asymptotics, let us discuss in some detail the evolution equation
for the mean, given by \fer{8mm}. This evolution equation has a
universal value, since it does not depend on the particular choice
of the map $\BigL$, but on its mean value only. Let $\e \ll 1$. Then, the scaling
 \be\label{8scal}
\b_1 \to \e\b_1,\quad \b_2 \to \e\b_2, \quad \mu \to \e\mu, \quad t \to \e
t = \tau,
 \ee
 is such that, while the elementary interaction is scaled by $\e$, the mean value of the density $f(v,t)$ satisfies
 \[
 \frac{dm_f(t)}{dt} = \e \b_2 m_f(t) + \e \mu e^{\e \b_1 t} = \e\left( \b_2 m_f(t) +  \mu e^{\b_1
 \tau} \right).
 \]
If we set $g_\e(v,\tau) = f(v,t)$, then $ m_{g_\e}(\tau) = m_f(t)$, and
the mean value of the density $g_\e(v,\tau)$ solves
  \be\label{8meanscal}
 \frac{dm_{g_\e}(\tau)}{d\tau} = \b_2 m_{g_\e}(\tau) +  \mu e^{\b_1
 \tau}.
 \ee
Note that equation \fer{8meanscal} does not depend explicitly on the
scaling parameter $\e$. In other words, the idea is to  reduce the growth of the
mutants by reducing the values of the constants, waiting enough time to get the same law for the mean value of the density of mutants. Note that the time scaling is equivalent to the scaling we did on Galton experiment in Section \ref{sec:Galton}, where we scaled the frequency $\tau \to \e$.

Using \fer{8scal}, we can write the kinetic equation \fer{8kine-w} in terms of
$g_\e(v,\tau)$. This equation reads
 \be
  \label{8kine-ww}
\frac{d }{d\tau}\int_{\R_+} \varphi(v) g_\e(v,\tau)\,dv = \frac
1\e\left\langle \int_{\R_+^2} \bigl( \varphi(v_\e^*)-\varphi(v)\bigr)
g_\e(v,\tau) M_\e(w,\tau) \,dv \, dw \right\rangle.
 \ee
In \fer{8kine-ww},
 \[
v_\e^*= v + \BigL_\e(v) + w,
 \]
and $\BigL_\e(v)$ denotes the map with mean growth $\e\b_2$.
Moreover, $M_\e(v,\tau)$ is the Poisson process with intensity
$\e\mu$ and growth parameter $\e\b_1$.

By means of equation \fer{8kine-ww} we can consequently investigate
the situation in which most of the interactions produce a very small
growth of mutants ($\e \to 0$), while at the same time  the
evolution of the density is such that \fer{8meanscal} remains
unchanged. In accord with the discussion of Section \ref{sec:Galton}, where a similar procedure has been applied to the kinetic model simulating Galton board, we will denote  this limit as the \emph{quasi-invariant growth} limit.

Under the scaling \fer{8scal}, equation \fer{8eq} for $g_\e(v,\tau)$ reads
\begin{equation}
  \label{8eq-s}
  \frac{\partial \GG_\e(\xi, \tau)}{\partial \tau}
= \frac 1\e \left[ \MM_\e(\xi,\tau)\int_{\R_+}\langle
e^{-i\xi\BigL_\e(v)}\rangle e^{-i\xi v} g_\e(v, \tau)\,dv  - \GG_\e(\xi,
\tau)\right].
 \end{equation}
 Let us observe that equation \fer{8eq-s} can be written equivalently
 as
 \be
  \label{8eq-s1}
  \frac{\partial \GG_\e(\xi, \tau)}{\partial \tau}
= \frac 1\e \int_{\R_+}\left( \langle e^{-i\xi\BigL_\e(v)}\rangle
\MM_\e(\xi,\tau) -1 \right) e^{-i\xi v} g_\e(v, \tau)\,dv,
 \ee
where
 \[
\MM_\e(\xi,\tau) = \exp\left\{\e\mu e^{\b_1 \tau}\left( e^{-i\xi} -
1\right) \right\}
 \]
is the Fourier transform of the Poisson process $M_\e(v,\tau)$. By
assuming that $\BigL(v)$ is a Poisson variable of mean value $\b_2 v$, so
that \fer{8pl} holds, \fer{8eq-s1} becomes
 \[
  \frac{\partial \GG_\e(\xi, \tau)}{\partial \tau}
= \frac 1\e \int_{\R_+}\left(\exp\left\{\e\left(\b_2v + \mu e^{\b_1
\tau}\right)\left(e^{-i\xi} -1 \right)\right\} -1 \right) e^{-i\xi v}
g_\e(v, \tau)\,dv,
 \]
which, expanding the exponential function in the integral in Taylor
series gives
 \bqa
 \nonumber
  \frac{\partial \GG_\e(\xi, \tau)}{\partial \tau}
&=& \left(e^{-i\xi} -1 \right)\int_{\R_+}\left(\b_2v + \mu e^{\b_1
\tau}\right) e^{-i\xi v} g_\e(v, \tau)\,dv + \e R(\xi, \tau)\\[-.25cm]
\label{8lim}
\\[-.25cm]
\nonumber
&=& \left(e^{-i\xi} -1 \right)\left( i\b_2 \frac{\partial  \GG_\e(\xi,
t)}{\partial \xi} + \mu e^{\b_1 \tau} \GG_\e(\xi, \tau) \right) + \e
R_\e(\xi, \tau) .
 \eqa
 The remainder term $R$ is such that
 \be\label{8rem}
|R_\e(\xi, \tau)| \le \frac 12 \vert e^{-i\xi} -1
\vert^2\int_{\R_+}\left|\b_2v + \mu e^{\b_1 \tau}\right|^2  g_\e(v,
\tau)\,dv .
 \ee
Standard computations then show that, for any fixed time $\tau >0$, the
remainder term remains uniformly bounded with respect to $\e$. Therefore,
letting $\e \to 0$ we obtain that the limit function $g(v,\tau)$ satisfies
the equation
  \be\label{8LC}
    \frac{\partial \GG(\xi, \tau)}{\partial \tau}
 = \left(e^{-i\xi} -1 \right)\left( i\b_2 \frac{\partial  \GG(\xi,
\tau)}{\partial \xi} + \mu e^{\b_1 \tau} \GG(\xi, \tau) \right).
 \ee

We remark that equation \fer{8LC} is the evolution equation for the
\index{cell mutation models!Lea--Coulson distribution} Fourier
transform of the Lea--Coulson distribution function \cite{Zhe0}.
Consequently, the Lea--Coulson formulation of the Luria--Delbr\"uck
distribution coincides with the quasi-invariant growth limit of the
kinetic model \fer{8eqLC}, where the self-growth of the mutated
cells follows a Poisson process of intensity proportional to the
number $v$ of mutated cells. Under this assumption, differently from
what happens in the original Luria--Delbr\"uck distribution, the
distribution function $g(v,\tau)$ takes values only on the positive
natural numbers.

Equation \fer{8LC} can easily be transformed back to the space of
probability distributions to get, for $n \ge 0$, the evolution
equation for the probabilities $g(n,\tau)$ of having $n$ mutant
cells at time $\tau$. These probabilities obey the recursive
equation
  \be\label{8LC1}
    \frac{\partial g(n, \tau)}{\partial \tau}
 = \b_2 \left( (n-1)g(n-1,\tau) - n g(n,\tau) \right) + \mu e^{\b_1 \tau} \left( g(n-1,\tau) -  g(n,\tau)
 \right),
 \ee
where $g(-1,\tau) = 0$. Note that, if we assume  that at time $\tau =0$ there are no mutants, the
initial conditions are given by
 \[
g(0, \tau = 0) = 1, \quad g(n, \tau = 0) = 0 \quad {\rm if} \quad n \ge 1.
 \]
Consider now that, for a given $\e >0$,  the Taylor expansion we
used to obtain formula \fer{8lim}  can be used to express equation
\fer{8LC} like a kinetic equation of type \fer{8eq-s1} plus a
remainder term which now depends on the solution $g(v,\tau)$ and its moments,
 \bqa
 \nonumber
    \frac{\partial \GG(\xi, \tau)}{\partial \tau}
 &=& \left(e^{-i\xi} -1 \right)\left( i\b_2 \frac{\partial  \GG(\xi,
\tau)}{\partial \xi} + \mu e^{\b_1 \tau} \GG(\xi, \tau) \right)\\[-.25cm]
\label{8LCappr}
\\[-.25cm]
\nonumber
&=&
  \frac 1\e \left[
\MM_\e(\xi,\tau)\int_{\R_+}\langle e^{-i\xi\BigL_\e(v)}\rangle e^{-i\xi v}
g(v, \tau)\,dv  - \GG(\xi, \tau)\right] - \e R(\xi, \tau).
 \eqa
The remainder term in \fer{8LCappr} takes the form
 \be\label{8rem1}
  R(\xi, \tau) =  \frac 12 \left(
e^{-i\xi} -1 \right)^2\int_{\R_+}\theta(\e,v)\left(\b_2v + \mu
e^{\b_1 \tau}\right)^2 e^{-i\xi v}g(v, \tau)\,dv ,
 \ee
 with $0 \le \theta(\e,v) \le 1$.
Since the mean and the variance of the Lea--Coulson distribution
remain bounded in time \cite{Zhe0}, formula \fer{8rem1} it holds
 \[
 \frac{R(\xi, \tau)}{|\xi|} \le \int_{\R_+}\left|\b_2v + \mu
e^{\b_1 \tau}\right|^2 g(v, \tau)\,dv = K(\tau) < +\infty.
 \]
Let now  $\GG$ and $\GG_\e$ denote the solutions to \fer{8eq-s} and \fer{8LC},
respectively, and let us define
 \[
h(\xi, \tau) = \frac{\GG(\xi, \tau) - \GG_\e(\xi, \tau)}{|\xi|}.
 \]
Using equation \fer{8LCappr}, and proceeding as in the computations
of formula \fer{8low}, we find that $h(\xi,\tau)$ satisfies
\begin{equation}
\partial_\t h (\xi,\tau) + \frac 1\e \,
h(\xi,\tau) \leq \frac{1+\e \b_2}\e \, \|h\|_\infty(\t) +\e K(\tau).
\end{equation}
This is equivalent to
 \[
 \partial_\tau \left (h(\xi,\tau) e^{\tau/\e} \right ) \leq
\frac{1+\e \b_2}\e \, \|h(\cdot, \tau) e^{\tau/\e}\|_\infty +
 \e K(\tau)e^{ \tau/\e}.
  \]
  Integrating from 0 to $\tau$, we
get
 \[
h(\xi,\tau) e^{\tau/\e} \leq h(\xi,0) + \int_0^\tau  \left ( \frac{1+\e
\b_2}\e \, \|h(\cdot, t) e^{t/\e}\|_\infty +
 \e K(t)e^{ t/\e}
 \right )\, dt.
 \]
 Hence, if $H_\e(\tau) =\|h(\cdot, \tau) e^{\tau/\e}\|_\infty$,
 \[
 H_\e(\tau) \leq H_\e(0) + \int_0^\tau \frac{1+\e
\b_2}\e \, H_\e(t) \, dt + \int_0^\t \e K(t)e^{ t/\e} \, dt.
 \]To the previous inequality we can apply the following classical generalized Gronwall inequality. If $u(\tau)$ satisfies
 \[
 u(\tau) \leq \phi(\tau) + \int_0^\tau \lambda(t) u(t)\, dt,
   \]
then it satisfies
 \[
  u(\tau) \leq \phi(0) \exp\left \{ \int_0^\tau \lambda(t)\, dt \right \}
+ \int_0^\tau \exp \left\{\int_t^\tau \lambda(t)\, dt \right \}
\frac{d\phi}{dt }\, dt.
 \]
Applying this inequality with $\lambda(\tau) = (1+\e \b_2)/\e$ and
$\phi(\tau) = H_\e(0) + \int_0^t \e K(t)e^{ t/\e} \, dt$, we obtain
 \[
 H_\e(\tau) \leq  \exp\left\{\frac{1+\e
\b_2}\e \right\}\left( H_\e(0) + \e \int_0^\tau K(t) e^{\b_2 t} dt
\right),
  \]
  namely
 \[
 \|h(\cdot, \tau)\|_\infty \leq e^{\b_2 \tau}\left( \|h(\cdot, 0)\|_\infty
 + \e \int_0^\tau K(t) e^{\b_2 t} dt \right).
 \]
Therefore, by choosing the same initial data for equations
\fer{8eq-s} and \fer{8LC}, so that $\|h(\cdot, 0)\|_\infty = 0$, we
obtain that, for $\tau
>0$,
 \[
 d_1( g, g_\e)(\tau) = \sup_{\xi} \frac{|\GG(\xi, \tau) - \GG_\e(\xi, \tau)|}{|\xi|}  \leq \e e^{\b_2 \tau}
\int_0^\t K(t) e^{\b_2 t} dt.
 \]
Letting $\e \to 0$ shows that the solution to the kinetic model
\fer{8eq-s} converges to the solution of the evolution equation for
the Lea--Coulson formulation \fer{8LC}. The previous computations
can be put in the form of a rigorous Theorem.
\begin{lemma}\label{8LCmain}
Let the probability density $f_0$ be defined as in \fer{8in}. Then,
choosing as parameters $\e\b_1$, $\e\b_2$ and $\e\mu$, as $\e \to
0$, the weak solution to the kinetic equation for the scaled density
$g_\e(v,\tau)=f(v,t)$, with $\tau = \e t$, converges to a
probability density $g(w,\tau)$. This density  is a weak solution of
the Fokker--Planck-like equation \fer{8LC}, whose solution is the
Fourier transform of the Luria--Delbr\"uck distribution in the
Lea--Coulson formulation.
\end{lemma}

The same strategy applies to the classical Luria--Delbr\"uck
distribution, \index{cell mutation models!Luria--Delbr\"uck
distribution} which appears as the weak limit of the kinetic
equation \fer{8eqLD}. In this case, using the same mathematical
tools, one can prove that the solution to \fer{8eqLD} converges to
the solution to the Fokker--Planck-like equation
  \be\label{8LD}
    \frac{\partial \GG(\xi, \tau)}{\partial \tau}
 = \b_2\xi \frac{\partial  \GG(\xi,
\tau)}{\partial \xi} + \mu e^{\b_1 \tau}\left(e^{-i\xi} -1 \right)
\GG(\xi, \tau) .
 \ee
Once again, the limit procedure leading to equation \fer{8LD}
illustrates the powerful role of the Fourier-based metric \fer{me-inf}.

\subsection{The Bartlett formulation}\index{cell mutation models!Bartlett model}
The kinetic approach is quite general, and can be extended to cover more general situations, including the Bartlett formulation \cite{Bar, Zhe0}. 

As extensively discussed in Section~\ref{sec:mutation}, in the
Lea--Coulson formulation of the Luria--Delbr\"uck distribution, the
growth of mutants is assumed to be a random process, while the
growth of normal cells is assumed to be completely deterministic. In
the Bartlett formulation the growth of both normal and mutant cells
is assumed to be fully stochastic. In this formulation
the Luria--Delbr\"uck model is represented by a two-dimensional
birth process $(X_1(t); X_2(t)), t\ge0 $, where $X_1(t)$ and
$X_2(t)$ represent the population size at time $t$ of the normal
cells and that of the mutant cells, respectively. The formulation is
called fully stochastic because it models the growth of both the
normal cells and the mutant cells by stochastic growth processes.

At a kinetic level, the Bartlett formulation belongs to a class of linear
kinetic equations for the distribution density $f = f(v,w,t)$ in which $v$
and $w$ represent the number of mutant and normal cells, respectively,
which are subject to the interaction rules
 \be\label{8ken}
v^* = v + \BigL_2(v) + w \, ; \quad w^* = w + \BigL_1(w).
 \ee
As for the simpler interaction \fer{8colli}, the two maps $\BigL_1$
and $\BigL_2$ can be either deterministic or random. According to
our standard assumptions,
 \[
\langle \BigL_1(w) \rangle = \b_1w \, ; \quad \langle \BigL_2(v) \rangle =
\b_2 v .
 \]
Resorting to the scaling of Section~\ref{8inva}, the effect of
interactions \fer{8ken} on the time variation of the density $f =
f(v,w,t)$ can be quantitatively described by a linear Boltzmann-type
equation, where the observable quantities $\varphi = \varphi(v,w)$ are now dependent on two variables. This equation reads
 \be
  \label{8kine-k}
\frac{d }{dt}\int_{\R_+^2} \varphi(v,w) f(v,w,t)\,dv\, dw = \langle
\int_{\R_+^2} \bigl( \varphi(v^*, w^*)-\varphi(v,w)\bigr) f(v,w,t)
\,dv \, dw \rangle.
 \ee
In \fer{8kine-k}, the presence of $\langle \cdot\rangle$ takes into
account the possibility that $\BigL_i(v), i =1,2$, could be  random
quantities. Note that, by choosing $\varphi(v,w) = 1$, one easily
recovers that $f(v,w,t)$ remains a probability density if it is so
initially,
 \be\label{8m00}
\int_{\R_+^2} f(v,w, t)\,dv \, dw= \int_{\R_+^2} f_0(v, w)\,dv\, dw = 1 .
 \ee
Moreover, on the basis of \fer{8ken}, choosing $\varphi(v, w) = v$
(respectively $\varphi(v, w) = w$) shows that the total mean numbers
$m_1(t)= \int_{\R_+^2} w f(v,w,t) $ and $m_2(t)= \int_{\R_+^2} v f(v,w,t)
$ of normal  and, respectively, mutant cells, satisfy the differential system
  \be\label{8me0}
 \frac{dm_1(t)}{dt} = \int_{\R_+^2}\langle \BigL_1(w)\rangle f(v,w, t)\,dv
 = \b_1 m_1(t),
  \ee
  \be\label{8mm3}
 \frac{dm_2(t)}{dt} = \int_{\R_+^2}(\langle \BigL_2(w)\rangle + w) f(v,w, t)\,dv
 = \b_2 m_2(t) + m_1(t).
  \ee
Note that, by solving \fer{8me0} with initial condition $m_1(0) =
\mu$ and substituting the result into \fer{8mm3}, one obtains that
the mean value $m_2(t)$ of mutants solves \fer{8mm}. Therefore, the
evolution of the mean of mutants in the Bartlett formulation
coincides with the evolution of the mean in the original
Luria--Delbr\"uck model.

Let us now suppose that the two maps $\BigL_i(v)$, $i=1,2$, are
independent random quantities. Then, by choosing $\varphi(v,w) =
e^{-i(\xi v + \eta w)}$, it is easily seen that the Fourier
transform of the solution to \fer{8kine-k},
 \[
\ff(\xi,\eta, t) = \int_{\R_+^2} e^{-i(\xi v + \eta w)}f(v,w, t)\,dv \,
dw,
 \]
satisfies the equation
\begin{equation}
  \label{8eqf}
  \frac{\partial \widehat f(\xi,\eta, t)}{\partial t}
  = \int_{\R_+^2 }\langle e^{-i\xi(\BigL_2(v)+w)}\rangle\langle e^{-i\eta\BigL_1(w)}\rangle
   e^{-i(\xi v+ \eta w)} f(v,w, t)\,dv  - \ff(\xi,\eta, t).
 \end{equation}
If now $\BigL_i(v)$, $i=1,2$ are Poisson distributed with means
$\b_i v, i =1,2$, so that
 \[
\langle e^{-i\xi\BigL_2(v)}\rangle = e^{\b_2
v\left(e^{-i\xi}-1\right)}, \quad \langle
e^{-i\eta\BigL_1(w)}\rangle = e^{\b_1 w \left( e^{-i\eta}
-1\right)},
 \]
equation \fer{8eqf} reduces to
 \be\label{8eqf2}
  \frac{\partial \ff(\xi,\eta, t)}{\partial t}
  = \ff \left( \xi + i\b_2
(e^{-i\xi}-1), \xi + \eta + i\b_1 (e^{-i\eta}-1), t \right) -
\ff(\xi,\eta, t),
 \ee
which represents the kinetic equation relative to the Bartlett formulation.
Note that the initial conditions on mass and mean value move to
 \be\label{8in-f}
\ff(0,0, t= 0) = 1, \quad \left.\frac{\partial \ff(\xi,\eta, t=
0)}{\partial \xi}\right|_{\xi = \eta = 0} = 0, \quad \left.\frac{\partial
\ff(\xi,\eta, t= 0)}{\partial \eta}\right|_{\xi = \eta = 0} = \mu.
 \ee
Let us apply to the parameters the same scaling as in \fer{8scal}.
Then proceeding as in Section~\ref{8inva} we find that the function
$\GG(\xi,\eta, \tau) = \ff(\xi,\eta, t)$, where $\tau = \e t$,
solves the equation
 \be\label{8eq-Ba}
  \frac{\partial \GG(\xi,\eta, \tau)}{\partial \tau}
  = i\b_2(e^{-i\xi}-1)\frac{\partial\GG}{\partial \xi} +\left( \xi +
i\b_1(e^{-i\xi}-1) \right)\frac{\partial\GG}{\partial \eta},
 \ee
with initial conditions given as in \fer{8in-f}. Equation
\fer{8eq-Ba} differs from equation $(95)$ in Zheng \cite{Zhe0} in
the coefficient of ${\partial\GG}/{\partial \eta}$. Reverting to the
physical space, we obtain, for $n,m \ge 0$, the evolution equation
for the probabilities $g(n,m,\tau)$ of having $n$ mutant cells and
$m$ normal cells at time $\tau$. These probabilities obey the
recursive equation
 \[
  \frac{\partial g(n,m, \tau)}{\partial \tau}
 = \b_2 \left( (n-1)g(n-1,m, \tau) - n g(n,m,\tau) \right) +
 \]
  \be\label{8LC2}
   \b_1 \left( (m-1)g(n,m-1, \tau) - m g(n,m,\tau) \right)
  \left( g(n-1,\tau) -  g(n,\tau)
 \right),
 \ee
where $g(-1,\tau) = 0$. If we assume, as in Section~\ref{sec:mutation},
that at time $\tau =0$ there are no mutants, the initial conditions
are given by
 \[
g(0, \tau = 0) = 1, \quad g(n, \tau = 0) = 0 \quad {\rm if} \quad n \ge 1.
 \]

\subsection{Numerical examples} \index{cell mutation models!Monte Carlo}
We compare the continuous distribution of mutants obtained using the
different kinetic models  in the generalized Fokker--Planck limit
and some standard methods for computing the approximated discrete
distribution in the Luria--Delbr\"uck  and Lea--Coulson settings
(see Lemma 2, page 11, in \cite{Zhe0}).

As usual, for the numerical solution of the kinetic models we can adopt
a Monte Carlo simulation method.  Here, the main difference
is the necessity of generating Poisson samples at each time step.
This can easily be achieved by standard algorithms; see, for example  
\cite{Kn,Mar}. For convenience of the reader we report the details of the Monte Carlo  
method used to solve the kinetic model \fer{8kine-w} under scaling \fer{8scal} in Algorithm \ref{alg:MC}.

\begin{algorithm}
\SetAlgoLined
\DontPrintSemicolon
 Define interaction parameters: $\beta_1$, $\beta_2$, $\mu\,;$\;
 Initialize algorithm parameters: $Ns$, $T_f$, $\varepsilon$, $n_T\,;$\;  
 Set initial conditions: $N_0=1$, $v_j=0$, $j=1,\ldots, N_s\,;$\;
 Set $t=0$, $\Delta t = T_f/(\e n_T)\,;$\;
 \For{k = 1, \dots, $n_T$}{
    $t = t + \Delta t$\,;\;
    $N_k=N_0e^{\e\beta_1 t}$\,;\;
    $M={\rm Sround}(\Delta t Ns)$\,;\;
    Select $M$ samples $v_h$ uniformly among the $Ns$ samples $v_j$ and for those compute\,:\;
    $\quad \eta_h \sim {\rm Poisson}(\e\mu N_{k-1})$\,;\;
    \quad For the Luria-Delbr\"uck case\,:\,
    $\mathcal{L}_h = \e\beta_2 v_h$\,;\;
    \quad For the Lea-Coulson case\,:
    $\mathcal{L}_h \sim {\rm Poisson}(\e\beta_2 v_h)$\,;\;
    $\quad v_h = v_h + \mathcal{L}_h+\eta_h$\,;\;
 }
 \caption{Monte Carlo method for the kinetic model \eqref{8kine-w} under scaling \fer{8scal}} 
 \label{alg:MC}
 
\end{algorithm}

Note that, in the above algorithm, the number of iterations $n_T$ scales as $1/\e$, since it has to be sufficiently large to guarantee that $\Delta t \leq 1$. The stochastic rounding ${\rm Sround}(x)$ rounds a real number $x$ to the next larger or smaller floating-point number with probabilities 1 minus the relative distances to those numbers.

The test cases considered were proposed in \cite{Zhe0}. We start
from initial conditions where no mutants are present:
$f_0(m)=\delta_0(m)$ and $N(0)=1$. The parameters are $\mu=10^{-7}$,
$\beta_1=3$ and the final computation time is $T_f=6.7$. To reduce
fluctuations, the total number of simulation samples $N_s$ is $5\times
10^5$. First we consider the Luria--Delbr\"uck case for $\beta_2=2.5$,
and then the Lea--Coulson case for $\beta_2=2.8$.

\begin{figure}[t]
\centering
\includegraphics[scale=0.35]{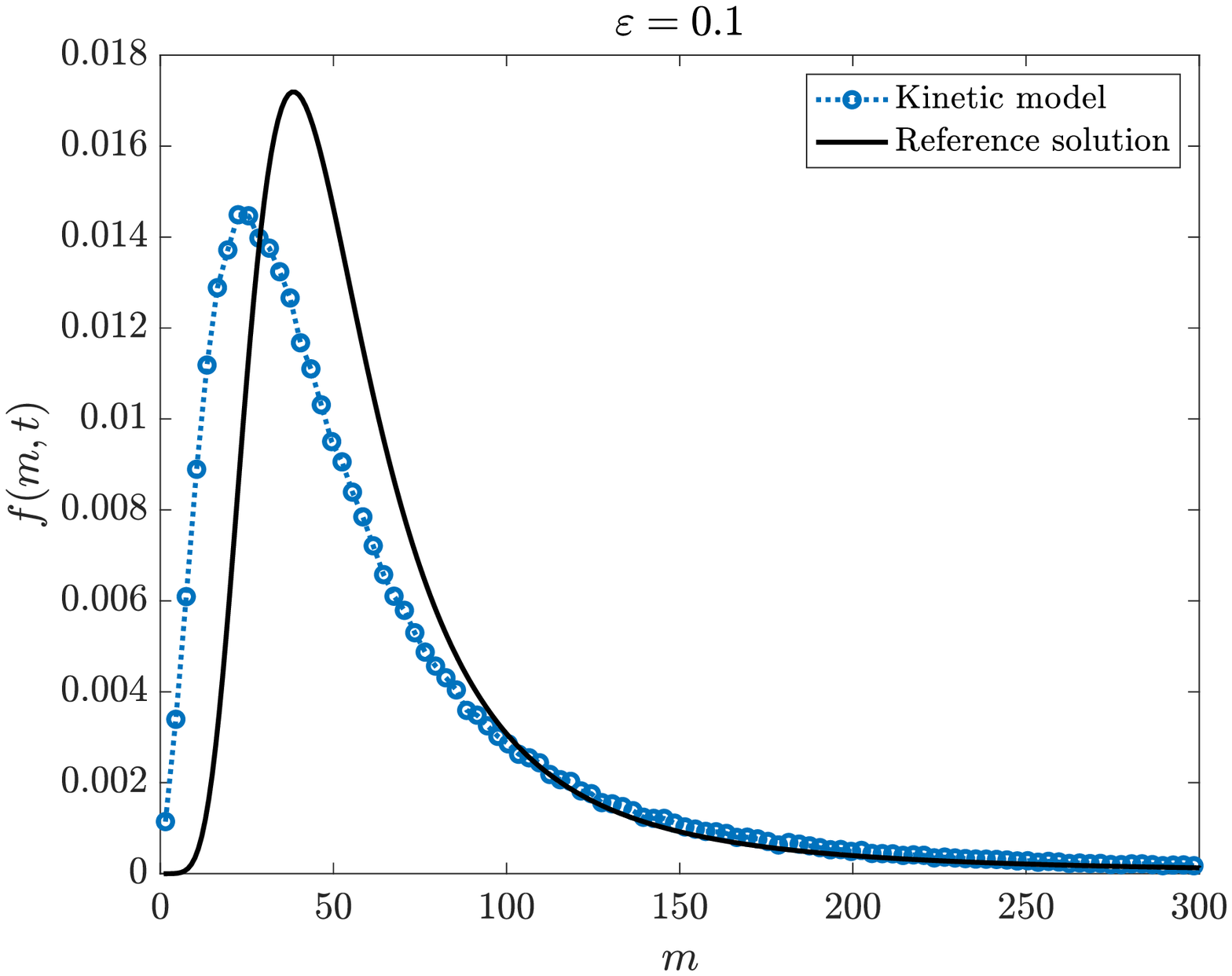}
\includegraphics[scale=0.35]{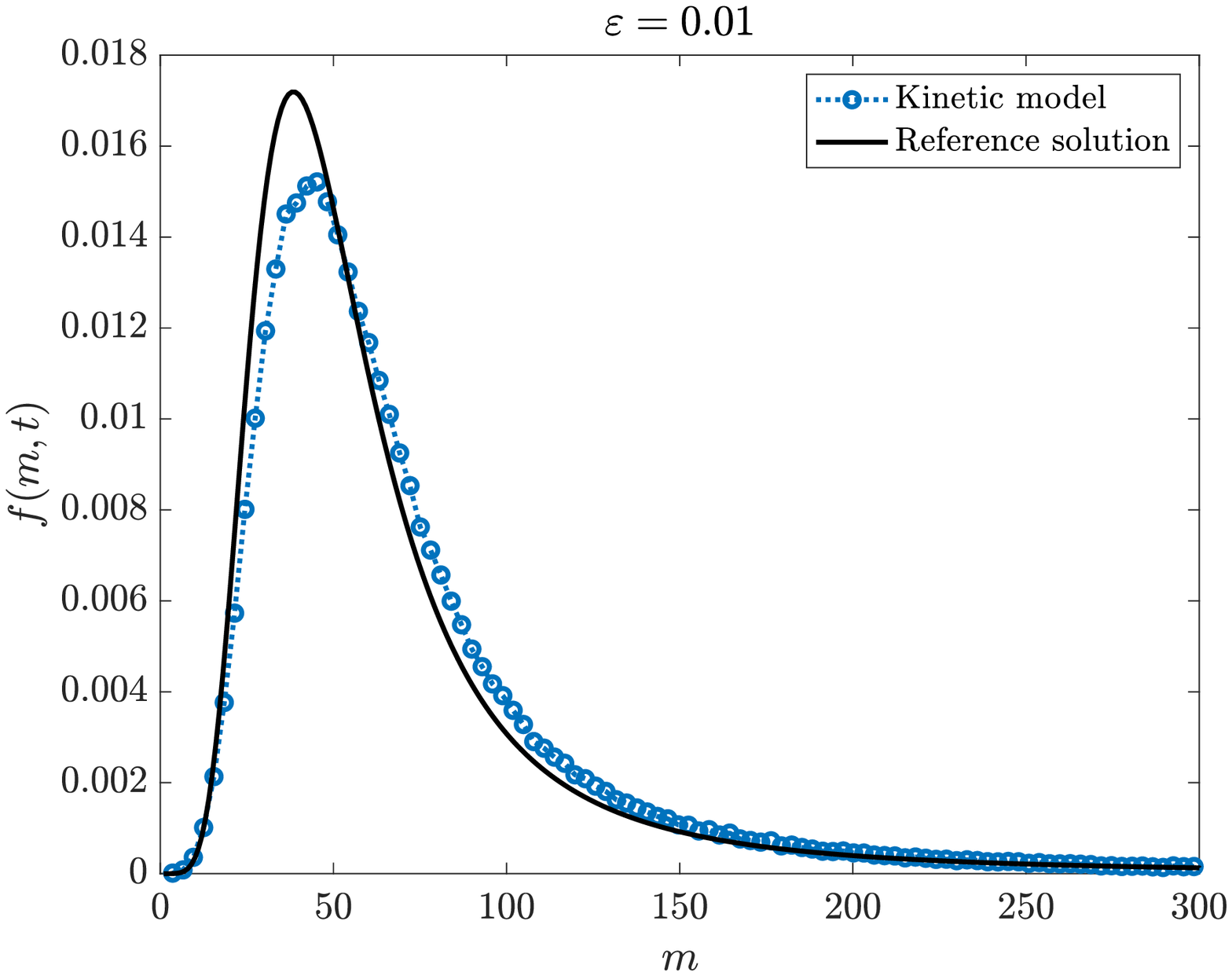}
\caption{Luria--Delbr\"uck case. Distribution of mutant cells at
$T_f=6.7$ with $\beta_1=3$, $\beta_2=2.5$ for $\varepsilon=0.1$ (left) and
$\varepsilon=0.01$ (right) in the kinetic model \fer{8kine-w} under scaling \fer{8scal}, with
$\BigL(v) = \beta v$. The reference solution is computed using Lemma
2 in \cite{Zhe0}.} \label{fg:k1}
\end{figure}

In Figure \ref{fg:k1} we report the solution obtained simulating the
kinetic model (\ref{8kine-ww}) for different values of the scaling
parameter $\varepsilon$, and for the map $\BigL_\e(v) = \e \beta_2 v$.
The results show the convergence of the mutant distribution
prescribed by the model towards the classical Luria--Delbr\"uck
solution computed as in \cite{Zhe0}.

\begin{figure}[t]
\centering
\includegraphics[scale=0.35]{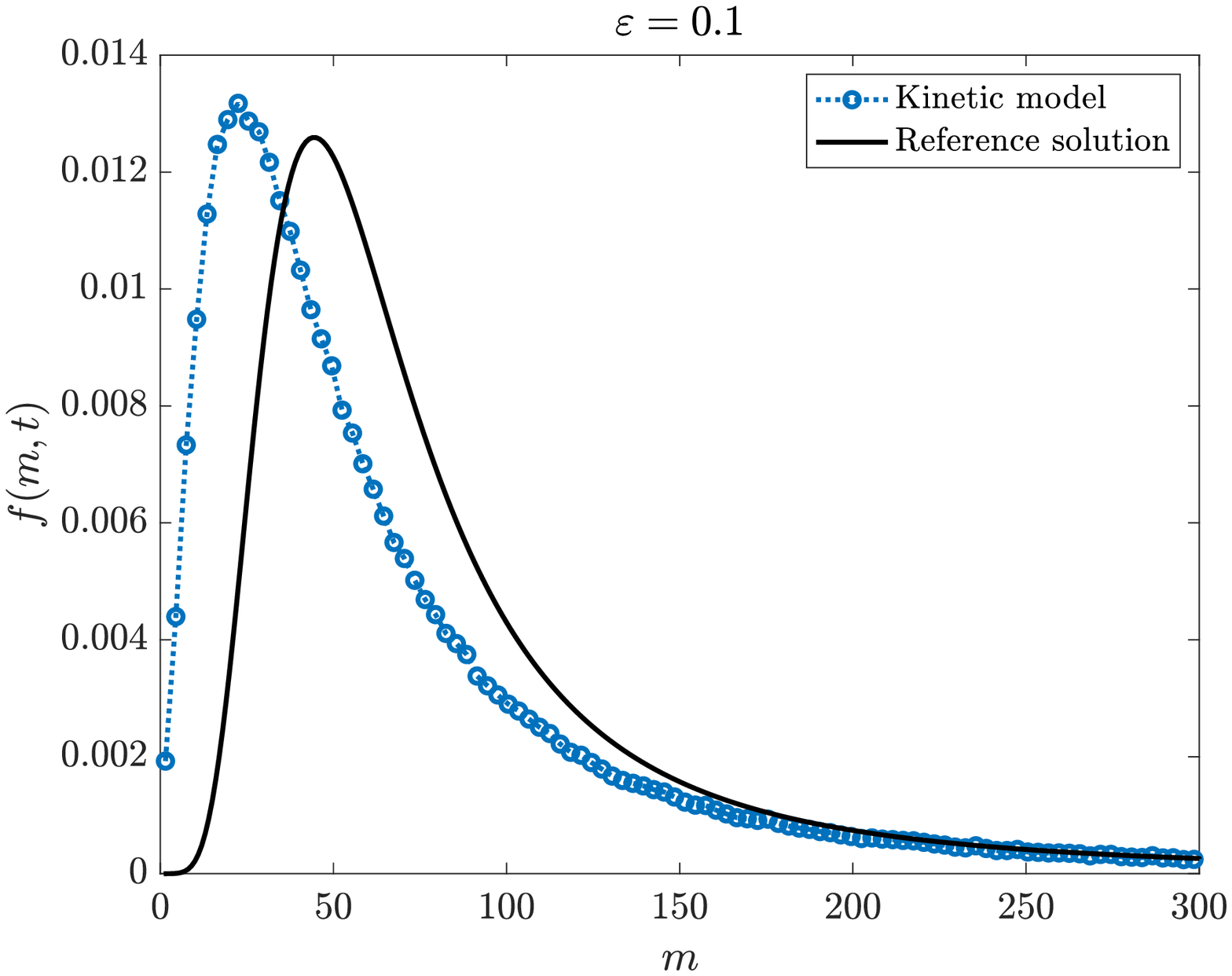}
\includegraphics[scale=0.35]{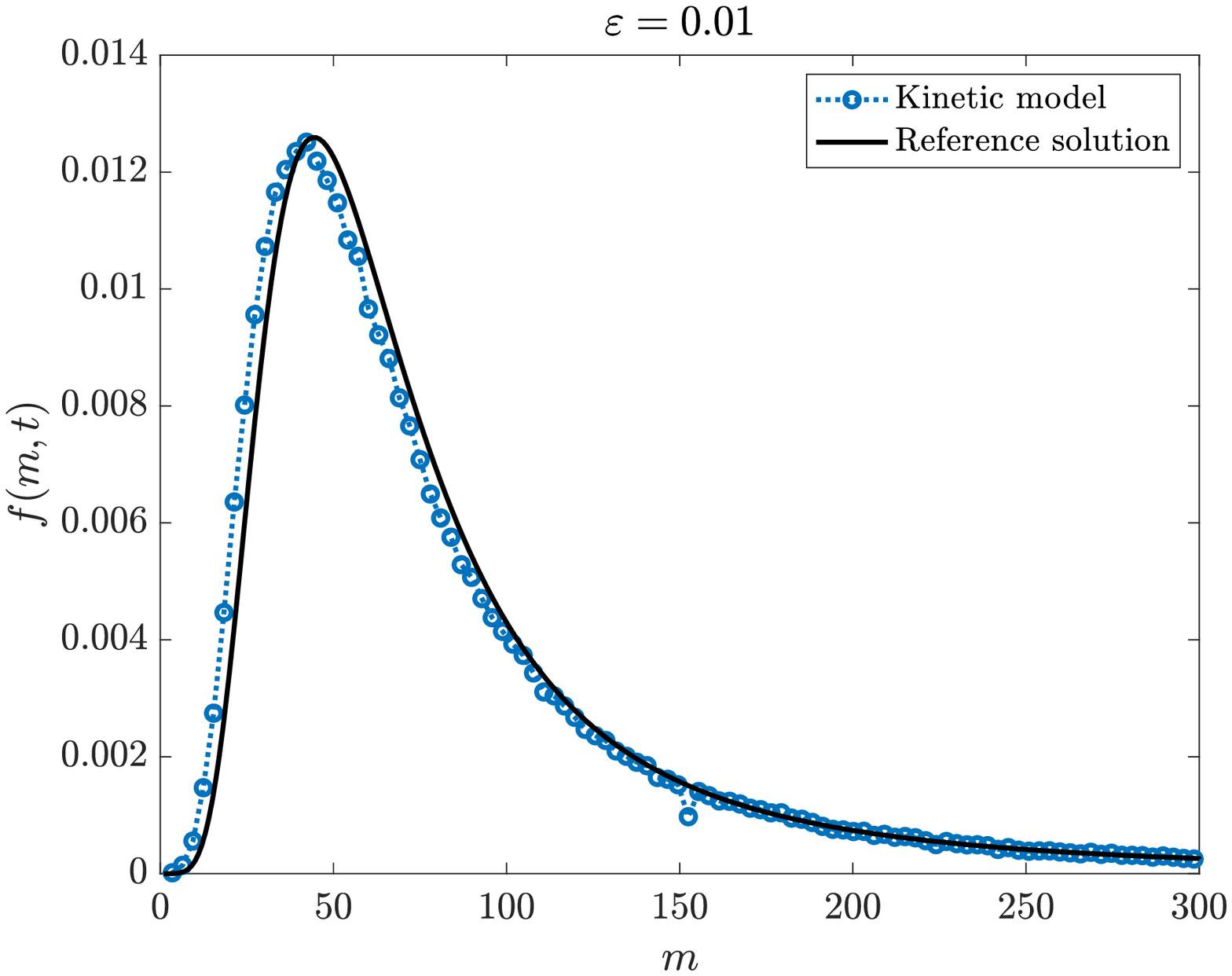}
\caption{Lea--Coulson case. Distribution of mutant cells at
$T_f=6.7$ with $\beta_1=3$, $\beta_2=2.8$ for $\varepsilon=0.1$ (left) and
$\varepsilon=0.01$ (right) in the kinetic model \fer{8kine-w} under scaling \fer{8scal}, with
$\BigL(v)$ a Poisson process of mean $\beta v$. The reference
solution is computed using Lemma 2 in \cite{Zhe0} and numerical
quadrature.} \label{fg:k2}
\end{figure}

Similarly, we report in Figure \ref{fg:k2} the solution of the
kinetic model (\ref{8kine-ww}) for different values of the scaling
parameter $\varepsilon$ and for the random map $\BigL_\e(v)$ such that
$\langle \BigL_\e(v)\rangle = \e \beta_2 v$. As expected,
convergence of the mutant distribution towards the Lea--Coulson
solution computed as in \cite{Zhe0} is observed.

\section{Conclusions}

Linear kinetic theory is a powerful tool for modeling  many particles systems subject to elementary interactions. Among the various problems that can benefit from this methodology, successful attempts have been made in recent years to study the classical problem of mutation rates, pioneered by the Luria and Delbr\"uck experiment. In this paper, following the work done in \cite{KP,To13}, we briefly presented the main building blocks that enable the construction of easy-to-handle mathematical models that use the relationships between elementary interactions to define the macroscopic behavior of the whole system. Theoretical results as well as numerical simulations show how these models can be used to characterize the asymptotic distribution of mutant cells.

\vspace{6pt} 





{\bf Acknowledgments:} {This work has been written within the activities of GNFM and GNCS groups of INdAM (National Institute of High Mathematics).  }






\end{document}